\documentclass[11pt]{article}

\usepackage{graphicx,amsmath,amsfonts,amssymb,dcolumn,amsthm,slashed,bbm,xspace,pgffor,tikz,mathbbol,latexsym,enumerate,amsmath}
\usepackage{soul}
\usepackage[colorlinks,citecolor=red,urlcolor=cyan,linkcolor=blue]{hyperref}
\usepackage[utf8]{inputenc}
\usepackage[english]{babel}
\usepackage{lmodern}
\usepackage{microtype}
\usepackage{cite}

\numberwithin{equation}{section}

\begin{document}

\title{\textbf{Generating Jackiw-Teitelboim Euclidean gravity from static three-dimensional Maxwell-Chern-Simons electromagnetism}}

\author{\textbf{Thales F.~Bittencourt}\thanks{thalesfernandesbittencourt@id.uff.br},~ \textbf{Rodrigo F.~Sobreiro}\thanks{rodrigo\_sobreiro@id.uff.br}\\
\textit{{\small UFF - Universidade Federal Fluminense, Instituto de F\'isica,}}\\
\textit{{\small Av. Litorânea, s/n, 24210-346, Niter\'oi, RJ, Brasil.}}}

\date{}
\maketitle

\begin{abstract}
We consider pure three-dimensional Maxwell-Chern-Simons electrodynamics in the static limit. We show that this theory can be mapped onto a two-dimensional gravitational model in the first-order formalism of Riemannian manifolds with Euclidean signature, coupled to a real scalar field naturally interpreted as a dilaton. In this framework, the Newtonian and cosmological constants in two dimensions are fully determined by the electric charge. The solution to this gravitational model is found to be trivial: a constant dilaton field on a flat manifold. However, we introduce two distinct shifts of the spin connection that transform the model into Jackiw-Teitelboim gravity. Specifically, we identify two additional solutions: a hyperbolic manifold with also a constant dilaton configuration; and a spherical manifold where, again, the dilaton assumes a constant, nonzero field configuration. In both non-flat cases, by employing the Gauss-Bonnet theorem in the specific cases of compact manifolds, we establish that the manifold's radius is fixed by the cosmological constant (and, therefore, by the electric charge).
\end{abstract}

\maketitle

\newpage

\tableofcontents

\newpage

\section{Introduction}\label{intro}

The primary objective of this work is to establish a connection between three-dimensional electrodynamics and Jackiw-Teitelboim (JT) gravity models \cite{Teitelboim:1983ux,Jackiw:1984je,Grumiller:2002nm,Maldacena:2019cbz,Nojiri:2000ja,Mertens:2022irh}, thereby contributing to the broader program of gauge/gravity correspondences. The well-known $AdS/CFT$ duality from string theory \cite{Maldacena:1997re} asserts that gravitational theories can often be described equivalently by gauge theories. The theories considered in this work are both simple and explicitly solvable, providing a concrete and accessible model through which this duality can be tested and understood in detail, without relying on the full machinery of string theory. Accordingly, the main system of interest in this paper is two-dimensional gravity \cite{Teitelboim:1983ux,Jackiw:1984je,Polyakov:1987zb,Gross:1989vs,Witten:1989ig,Chamseddine:1989dm,Fukuma:1990jw,Kazama:1992ex,Grumiller:2002nm,Mertens:2022irh,Sybesma:2022nby}. In particular, two-dimensional gravity has served as a robust framework for probing the quantization of gravity in higher dimensions. Its topological character further motivates the investigation of topological field theories as candidates for quantum gravity in four dimensions (see, for instance, \cite{Witten:1989ig,Alvarez:2011gd,Junqueira:2021rwc,Sadovski:2024uhg} and references therein). Another compelling reason for studying gravity theories across different dimensions is that numerous physical systems can be effectively described by geometrodynamical models \cite{Witten:1988hc,Obukhov:1998gx,Sobreiro:2007pn,Sobreiro:2011hb,Assimos:2013eua,Assimos:2019yln,Assimos:2021eok}. In such models, gravity (or geometrodynamics) provides an alternative framework for describing certain gauge theories. 

Specifically, in the case of JT gravity, numerous potential applications and significant results have been investigated \cite{Isler:1989hq,Cadoni:1994uf,Brown:2018bms,Moitra:2019bub,Almheiri:2019qdq,Almheiri:2019psf,Maldacena:2019cbz,Iliesiu:2019xuh,Nojiri:2000ja,Mertens:2022irh,Nejati:2023hpe,Pinzul:2024zkl,Dowd:2024gef}. It plays a particularly important role in black hole physics. While the Einstein-Hilbert model with cosmological constant in two dimensions simply defines a Plateau problem \cite{Mardones:1990qc,Zanelli:2005sa}, the JT model exhibits richer dynamical features due to the presence of the dilaton, making it a good description for two-dimensional black holes and for a large group of near-extremal black holes close to their horizons in higher dimensions \cite{Mertens:2022irh,Cadoni:1994uf,Nayak:2018qej,Moitra:2019bub,Nojiri:2000ja}. Thus, establishing a correspondence with Maxwell–Chern–Simons electrodynamics suggests that thermodynamic and quantum features of black holes — such as entropy and near-extremal dynamics — might have analogues in simple electromagnetic gauge systems. Another major area where JT gravity has had a strong impact is holography (see \cite{Ecker:2021guy} and references therein). Typically, JT models have constant and negative curvature as solutions, stating an $AdS_2$ spacetime (two-dimensional Anti-de Sitter spacetime). In this manner, $AdS/CFT$ correspondence \cite{Maldacena:1997re} plays a central role in studying quantum gravity from a string-theoretic perspective. By analyzing such a lower-dimensional system, the underlying structure of quantum gravity may emerge more transparently, providing valuable insights for higher dimensions. In addition, JT gravity is also relevant in matrix models \cite{Stanford:2019vob,Witten:2020wvy}, which are closely related to Sachdev–Ye–Kitaev (SYK) systems \cite{Sachdev:1992fk,Kitaev2015} to describe chaotic quantum systems in condensed matter physics. It turns out that the SYK model is deeply connected to black hole physics due to a duality to JT gravity via $AdS/CFT$ correspondence (see, for example, \cite{Forste:2017apw,Nojiri:2000ja} and references therein).

The second system considered in this work is three-dimensional electrodynamics, which allows for the inclusion of the Chern-Simons (CS) topological term. Electromagnetic models of this kind have applications in various branches of physics, extending beyond purely field-theoretical investigations. Notably, they have been employed to describe aspects of topological insulators \cite{Santos:2015raa,Martin-Ruiz:2015skg}, Weyl semimetals and axion effects \cite{Goswami:2012db,Gorbar:2018vuh}, graphene layers \cite{Dudal:2018mms,Frassdorf:2017zqt,Capozziello:2018mqy,Carrington:2019rlt}, and even three-dimensional gravity \cite{Achucarro:1987vz,Witten:1988hc,Zanelli:2012px,Assimos:2019yln,Assimos:2021eok}. Moreover, the consistency of CS theories at the quantum level is a highly valued feature. In fact, quantum CS theories have long been known to be perturbatively finite \cite{Blasi:1989mw,Piguet:1995er}. Relating this theory to JT gravity thus enables a reinterpretation of certain topological or confinement features of the gauge system in purely geometric terms, and, conversely, allows field-theoretical tools to be applied to gravitational problems. Additionally, the existence of a mapping between these theories provides an exactly solvable framework to test ideas about gauge/gravity duality, offers a fresh perspective on the quantization of Chern–Simons theories, and opens the possibility of employing condensed matter systems as real laboratory simulators of gravitational dynamics.

The approach we adopt to relate electrodynamics to gravity theories was previously explored in two recent studies by the authors \cite{Sobreiro:2021leg,Bittencourt:2025laf}. In those works, we showed that two-dimensional gravity can emerge from two-dimensional electrodynamics (including fermions) by defining a suitable mapping between the fields of both theories. This mapping arises from the isomorphism between the groups $U(1)$ and $SO(2)$. It is important to emphasize that the mapping is rigorously defined in Euclidean spaces — i.e., electrodynamics formulated on $\mathbb{R}^2$ spacetime is mapped onto gravity on a two-dimensional manifold equipped with a Euclidean signature metric. As will become evident throughout the paper, this feature is particularly advantageous in the present study. 

At this point, it is also worth mentioning that the gravity models considered here are formulated within the first-order formalism (FOF) \cite{Utiyama:1956sy,Kibble:1961ba,Sciama:1964wt,Mardones:1990qc,Zanelli:2005sa}. The FOF is a theoretical framework in which gauge symmetry is identified with local spacetime isometries, thereby connecting gravity more closely to standard gauge theories. Moreover, in this formalism, the fundamental fields are the vielbein (or zweibein in two dimensions) and the spin connection, rather than the metric tensor and affine connection used in the Palatini formalism \cite{Olmo:2011uz}, although the correspondence between the two formalisms is well defined \cite{Sadovski:2022kwf}. Additionally, we assume metric compatibility and vanishing torsion. Note that in the FOF of gravity the gauge curvature is also the spacetime curvature. On the other hand, electrodynamics with the CS term is a gauge theory directly tied to topological invariants, but with a gauge curvature that has no relation to spacetime curvature since its gauge symmetry is an internal one. Establishing a correspondence between both theories would enable one to rephrase topological data in a gauge theory of internal symmetry straightforwardly as geometry, thus bridging topological and geometric approaches. 

We begin this work by considering a three-dimensional electrodynamics action comprising Maxwell and CS terms, \emph{i.e.}, the Maxwell-Chern-Simons (MCS) theory. This theory is then dimensionally reduced by imposing the static limit, leading to a two-dimensional magnetic theory with an additional scalar field. After exploring the fundamental aspects of this system, we proceed to map it onto a gravity theory, following the formalism developed in \cite{Sobreiro:2021leg,Bittencourt:2025laf}. In this context, the gravitational constants (Newton’s constant and the cosmological constant) are entirely determined by the electric charge. Since the electric charge is known \cite{ParticleDataGroup:2024cfk}, it can be used to obtain explicit values for these gravitational constants. Interestingly, we find that Newton’s constant and the cosmological constant are significantly larger when compared to their four-dimensional experimental values \cite{ParticleDataGroup:2024cfk}. Specifically, Newton’s constant is found to be $0.22eV^{-1}$, while the cosmological constant is valued as $0.09eV$. 

Furthermore, we identify a two-dimensional Euclidean dilatonic gravity model whose solution is trivial: the dilaton field remains constant throughout the manifold, while the manifold itself is flat with minimal area. Nevertheless, we demonstrate that the model can be reformulated to describe JT gravity with two distinct nontrivial solutions. The first corresponds to a manifold with a negative constant curvature and a constant dilaton, \emph{i.e.}, the model admits hyperbolic spaces as solutions \cite{Stillwell1992,Anderson2005}. The second non-flat solution describes a regular spherical manifold again with a constant dilaton configuration. 

In the three cases (flat, hyperbolic, and spherical), the Gauss-Bonnet (GB) theorem can be employed to determine additional properties of these manifolds. Particularly, for compact surfaces, we can study: the flat torus; the $n$-holed torus (negative curvature); and the sphere (positive curvature). Except for the flat case, we are able to determine the characteristic length scale of such surfaces, which is found to be on the order of $\mu m$ (micrometers). This is possible because the GB theorem relates the area of surfaces to their Euler characteristics \cite{Nakahara:2003nw,Stillwell1992}, and because the area depends on the cosmological constant (explicitly calculated from the electric charge). In addition, it is crucial to emphasize that the CS term plays a fundamental role in achieving the JT model. In fact, within our framework, the JT term is directly proportional to the CS level of the theory. 

This work is organized as follows. In Section \ref{3DQED}, we introduce three-dimensional MCS electrodynamics, discuss its main features, and consider its static limit. In Section \ref{GRAV} we map the static electrodynamics onto a two-dimensional gravity theory in the FOF and analyze its flat solution. Then, in Section \ref{JT} we relate this gravity theory to JT gravities and examine the main properties of these solutions. In Section \ref{GB}, we compute, using topological arguments, the radius of some manifolds, specifically for the $n$-holed tori and the sphere. Finally, our conclusions and perspectives are presented in Section \ref{conc}.

\section{Three-dimensional electrodynamics}\label{3DQED}

In this section, we define three-dimensional MCS electrodynamics and, by imposing the static limit, we dimensionally reduce it to two spatial dimensions.

 \subsection{The Maxwell-Chern-Simons theory}

The starting point is the most general three-dimensional action for pure electrodynamics that is polynomial in the fields and their derivatives and is power-counting renormalizable. The resulting action is simply the pure MCS action,
\begin{equation}
    S_{3qed}=\int\left(-\frac{1}{\mathrm{e}^2}f\underline{\ast} f+\kappa af\right)\;,\label{sqed1}
\end{equation}
where $a=a_\mu dx^\mu$ denotes the gauge field, $f=\underline{d}a$ denotes the field strength, and $\underline{\ast}$ represents the three-dimensional Hodge dual. The notation $\underline{d}=\partial_\mu dx^\mu$ stands for the three-dimensional exterior derivative. Greek indices run over $\{0,1,2\}$ with the $0$th coordinate characterizing time. Moreover, we consider a three-dimensional Minkowski spacetime with a negative signature as the background for MCS theory \eqref{sqed1}. We also note that natural units are used throughout.

The first term in \eqref{sqed1} is clearly the Maxwell action; therefore, $\mathrm{e}$ is the electric charge. The second term in \eqref{sqed1} is the CS topological term and $\kappa$ denotes the CS level of the theory. The canonical dimensions of fields and parameters are given by $\left[a\right]=1$, $\left[\mathrm{e}\right]=1/2$, and $\left[\kappa\right]=0$. The field equations can be derived straightforwardly from the action \eqref{sqed1} by applying Hamilton's principle, yielding
\begin{eqnarray}
    \underline{d}\underline{\ast}f+\kappa \mathrm{e}^2 f=0\;,\label{feqqed1}
\end{eqnarray}
Moreover, the Bianchi identities also hold, namely,
\begin{eqnarray}
    \underline{d}f=0\;.\label{feqqed2}
\end{eqnarray}
By performing a suitable and straightforward spacetime decomposition, one can easily infer that the field strength $f$ is related to the electromagnetic fields as
\begin{equation}
    f^{\mu\nu}\equiv\begin{pmatrix}
0 & -E^1 & -E^2\\
E^1 & 0 & B\\
E^2 & -B & 0
\end{pmatrix}\;.\label{fstr1}
\end{equation}
This field strength describes an electrodynamics governed by a two-dimensional electric vector field and a pseudoscalar magnetic field, as expected. The equations of motion can be decoupled in the standard way to obtain massive wave equations,
\begin{eqnarray}
    \left(\Box-m^2\right){\bf E}&=&0\;,\nonumber\\
    \left(\Box-m^2\right)B&=&0\;,\label{wave1}
\end{eqnarray}
with $\Box=\partial_t^2-\nabla^2$ denoting the d’Alembertian operator, ${\bf E}=(E^1,E^2)$, and $m=\kappa \mathrm{e}^2$ the CS topological mass. At the classical level, these waves decay exponentially over a characteristic length scale of order $m^{-1}$. This is the standard result of MCS electromagnetism.

Finally, it is worth recalling that electrodynamics is known to be a gauge theory invariant under $U(1)$ gauge transformations,
\begin{equation}
    \delta a=\underline{d}\alpha\;,\label{gt1}
\end{equation}
with $\alpha$ being a local parameter. In fact, electrodynamics can be fundamentally formulated as a $U(1)$ gauge theory \cite{Rubakov:2002fi}. It turns out that the $U(1)$ symmetry is the key for the mapping connecting this theory to gravity.

\subsection{Dimensional reduction - The static limit}

Once the MCS theory is defined, dimensional reduction of the model can be performed. We choose to reduce the time coordinate by considering only static systems, which provides the simplest way to eliminate the time dimension. Therefore, decomposing space and time of the exterior derivative, the gauge field, and field strength, we obtain
\begin{eqnarray}
    \underline{d}&=&\partial_tdt+d\;,\nonumber\\
    a&=&\Phi dt+A\;,\nonumber\\
    f&=&\left(d\Phi-\partial_tA\right)dt+F\;,\nonumber\\
    \underline{\ast}f&=&-\ast\left(d\Phi-\partial_tA\right)+\ast Fdt\;,\label{decomp1}
\end{eqnarray}
with $A=A_idx^i$ denotes the two-dimensional gauge field, $\Phi$ denotes the electric potential, and $F=dA$ is the two-dimensional electromagnetic field strength. Here, $\ast$ is the Hodge dual in the two-dimensional Euclidean space while $d=\partial_idx^i$ is the purely spatial exterior derivative. The Latin indices $i,j,k\ldots$ run over $\{1,2\}$, and describe the Euclidean spatial sector. Accordingly, the Lagrangian terms decompose as
\begin{eqnarray}
    f\underline{\ast}f&=&\left[\left(d\Phi-\partial_tA\right)\ast\left(d\Phi-\partial^tA\right)+F\ast F\right]dt\;,\nonumber\\
    af&=&\left(\Phi F+Ad\Phi-A\partial_tA\right)dt\;.\label{decomp2}
\end{eqnarray}
The corresponding decomposition of the action \eqref{sqed1} is just
\begin{equation}
    S_{3qed}=\int\left\{-\frac{1}{\mathrm{e}^2}\left[\left(d\Phi-\partial_tA\right)\ast\left(d\Phi-\partial^tA\right)+F\ast F\right]+\kappa\left[\Phi F+A\left(d\Phi-\partial_tA\right)\right]\right\}dt\;.\label{sqed2}
\end{equation}

Dimensional reduction can be carried out by restricting attention to static systems. In this case, all time derivatives vanish, while the factor $\int dt$ appears only as a normalization constant to be removed. The action \eqref{sqed2} then reduces to the static electrodynamical action
\begin{equation}
    S_{3stqed}=\int\left[-\frac{1}{\mathrm{e}^2}\left(d\Phi\ast d\Phi+F\ast F\right)+\kappa\left(\Phi F+Ad\Phi\right)\right]\;,\label{sqed3}
\end{equation}
where integration is taken over two-dimensional Euclidean space. Moreover, since $Ad\Phi=\Phi F-d(\Phi A)$, we may rewrite action \eqref{sqed3} as
\begin{equation}
    S_{3stqed}=\int\left\{-\frac{1}{\mathrm{e}^2}\left(d\Phi\ast d\Phi+F\ast F\right)+\kappa\left[2\Phi F-d\left(\Phi A\right)\right]\right\}\;,\label{sqed4}
\end{equation}
so that the term $Ad\Phi$ reduces to a surface contribution. This boundary term can, of course, play a role in systems with non-trivial boundaries, such as topological insulators or related setups.

The action \eqref{sqed4} can be regarded as a form of three-dimensional static electromagnetism, which admits different interpretations. For example, it can be viewed as electrodynamics in two Euclidean dimensions coupled to a scalar field. Equivalently, from the two-dimensional perspective, the static model may be seen as a scalar field $\Phi$ interacting with a pseudo-scalar magnetic field through the reduced field strength
\begin{equation}
    F^{ij}\equiv\begin{pmatrix}
 0 & B\\ 
 -B & 0
\end{pmatrix}\;.\label{fstr2}
\end{equation}

It is worth noting that the present situation differs essentially from the previous works \cite{Sobreiro:2021leg,Bittencourt:2025laf} where the two-dimensional electrodynamics under consideration was purely electric. A purely electric system can be obtained by reducing the third dimension ($\mu=2$) instead of the time coordinate. However, the resulting electrical model would still reside in Minkowskian spacetime. Such a model could not be mapped in a gravity model in an easy way as in the Euclidean case. This is the main reason why we perform dimensional reduction along the time direction rather than a spatial one. See \cite{Sobreiro:2021leg,Bittencourt:2025laf} for further details.

The static field equations can be obtained directly from the field equations \eqref{feqqed1} or from the minimization of the action \eqref{sqed4}. The equations for the scalar and vector fields are
\begin{eqnarray}
    d\ast d\Phi+mF&=&0\;.\nonumber\\
    d\ast F+md\Phi&=&0\;.\label{feqqed3}
\end{eqnarray}
These equations can be decoupled into
\begin{eqnarray}
    \left(\nabla^2-m^2\right)\Phi&=&C\;,\nonumber\\
    \left(\nabla^2-m^2\right)A&=&G\;,\label{feqqed5}
\end{eqnarray}
where $G$ and $C=C_idx^i$ are integration constants that may, without loss of generality, be set to zero. These are massive Laplace equations whose solutions, in polar coordinates, are expressed in terms of the modified Bessel functions of the first and second kinds (depending on $mr$, with $r$ being the radial coordinate), and periodic functions in the angular coordinate.

It is also useful, for future purposes, to note that after the dimensional reduction, the gauge transformations \eqref{gt1} decouple to
\begin{eqnarray}
    \Phi'&=&\Phi\;,\nonumber\\
    A'&=&A+d\alpha\;,\label{gt2}
\end{eqnarray}
\emph{i.e.}, the scalar field becomes invariant under gauge transformations, while the magnetic vector potential remains as a gauge field.

As an extra and final step, we rewrite the action \eqref{sqed4} with the aid of an auxiliary field (see \cite{Obukhov:1998gx,Sobreiro:2021leg,Bittencourt:2025laf}),
\begin{equation}
    S_{3qed}=\int\left\{-\frac{1}{\mathrm{e}^2}\left(d\Phi\ast d\Phi+\Theta\ast F-\frac{1}{2}\Theta\ast \Theta\right)+\kappa\left[2\Phi F-d\left(\Phi A\right)\right]\right\}\;.\label{sqed5}
\end{equation}
The field equation for the auxiliary field $\Theta$ trivially reads $\Theta=F$, which reproduces the action \eqref{sqed4} again. Naturally, the introduction of auxiliary fields in this manner is also valid at the quantum level \cite{Itzykson:1980rh}.

Finally, for completeness, the canonical dimensions and form ranks of the fields and parameters appearing in the action \eqref{sqed5} are displayed in Table \ref{table0a}.
\begin{table}[ht]
\centering
\begin{tabular}{|c|c|c|c|c|c|}
	\hline 
Fields & $A$ & $\Theta$ & $\Phi$ & $\mathrm{e}$ & $\kappa$ \\
	\hline 
Dimension & $1$ & $2$ & $1$ & $1/2$ & $0$ \\ 
Rank & $1$ & $2$ & $0$ & $0$ & $0$ \\
\hline 
\end{tabular}
\caption{Canonical dimensions and form ranks of the electromagnetic fields and parameters.}
\label{table0a}
\end{table}

\section{Achieving gravity and the flat solution}\label{GRAV}

In this section, we map the static electromagnetic action \eqref{sqed5} onto a two-dimensional gravity theory defined on manifolds with Euclidean signature. The technical details and formal aspects of the mapping can be found in \cite{Sobreiro:2021leg,Bittencourt:2025laf}. In those works, extra auxiliary non-physical fields are introduced in order to make the mapping well defined. Here, we omit such particularities because the same method applies in the present case. In short, the mapping is based on the isomorphism between the gauge group $U(1)$ of electrodynamics and the $SO(2)$ rotations of two-dimensional Euclidean gravity. The map between the fields of electrodynamics and gravity is given by\footnote{The factor $\mathrm{e}^4$ is different from the pure two-dimensional case \cite{Sobreiro:2021leg,Bittencourt:2025laf} because the canonical dimension of the electric charge is different. In fact, if $D$ is spacetime dimension, the canonical dimension of the electric charge is $[\mathrm{e}]=(4-D)/2$.}
\begin{eqnarray}
\Theta(x)&\longmapsto&\mathrm{e}^4\epsilon_{ab}e^a(X)e^b(X)\;,\nonumber\\
A(x)&\longmapsto&\epsilon_{ab}\omega^{ab}(X)\;,\nonumber\\
\Phi(x)&\longmapsto&\Phi(X)\;,\label{map1}
\end{eqnarray}
with $e^a=e^a_idX^i$ being the zweibein 1-form, with vanishing mass dimension $[e]=0$, while $\omega^{ab}=\omega^{ab}_idX^i$ is the spin connection, with mas dimension given by $[\omega]=1$. The point $x\in\mathbb{R}^2$ belongs to the Euclidean space where action \eqref{sqed5} is defined and $X\in M^2$ is a point in the two-dimensional manifold $M^2$ where gravity will emerge. The indices $a,b,c,\ldots,h\in\{1,2\}$ are associated with the tangent space of $M^2$, essentially a local inertial frame, while $i,j,k,\ldots\in\{1,2\}$ refer to a general coordinate system of the manifold $M^2$. The core idea of the mapping is the identification of the fields of three-dimensional static electromagnetism with the geometric fields of the manifold; in other words, the dynamics of the electromagnetic fields are absorbed into the manifold.

Performing the mapping \eqref{map1} in the action \eqref{sqed5}, one straightforwardly obtains
\begin{equation}
    S_{2grav}=\int\left\{-\frac{1}{\mathrm{e}^2}\left(d\Phi\ast d\Phi+2\mathrm{e}^4\epsilon_{ab}R^{ab}-\mathrm{e}^8\epsilon_{ab}e^ae^b\right)+\kappa\left[2\Phi\epsilon_{ab}R^{ab}-\epsilon_{ab}d\left(\Phi\omega^{ab}\right)\right]\right\}\;,\label{smap1}
\end{equation}
where $R^{ab}=d\omega^{ab}$ is the curvature 2-form and $\epsilon_{ab}$ being the Levi-Civita symbol. Therefore, by inferring the parameter identifications\footnote{Obviously, $[G]=-1$ and $[\Lambda]=1$.},
\begin{eqnarray}
    G&=&\frac{1}{16\pi \mathrm{e}^2}\;,\nonumber\\
    \Lambda&=&\mathrm{e}^2\;,\label{id0}
\end{eqnarray}
we find the gravity model,
\begin{equation}
    S_{2grav}=\int\left\{-\frac{1}{\Lambda}d\Phi\ast d\Phi-\frac{1}{8\pi G}\epsilon_{ab}\left(R^{ab}-\frac{\Lambda^2}{2}e^ae^b\right)+\kappa\left[2\Phi\epsilon_{ab}R^{ab}-\epsilon_{ab}d\left(\Phi\omega^{ab}\right)\right]\right\}\;.\label{sgrav1}
\end{equation}
This is a two-dimensional gravity action, in the first-order formalism, coupled to a scalar field naturally identified as the dilaton. The constants $G$ and $\Lambda$ are interpreted as the two-dimensional Newtonian and cosmological constants, respectively. Remarkably, since the electric charge is known, the numerical values of these gravitational constants can be computed through expressions \eqref{id0}. Indeed, 
\begin{eqnarray}
    G&\approx&0.22eV^{-1}\;,\nonumber\\
    \Lambda&\approx&0.09eV\;.\label{num0}
\end{eqnarray}
Hence, $G$ and $\Lambda$ are very large if compared with their four-dimensional physical values \cite{ParticleDataGroup:2024cfk}, namely $G\approx6.7\times10^{-57}eV^{-2}$ and $\Lambda\approx2.1\times10^{-33}eV$. Moreover, from \eqref{id0}, these constants are constrained by
\begin{equation}
    G\Lambda=\frac{1}{16\pi}\;.\label{id1}
\end{equation}

The first term in the action \eqref{sgrav1} is the standard kinematic term for a real scalar field, up to a constant factor. The second and third terms are the typical Einstein-Hilbert (EH) and cosmological constant terms. The EH action, in two dimensions, is known to be topological, a property reflected by the fact that it reduces to a total derivative. In fact, in two dimensions, the EH term coincides with the GB term. See Section \ref{GB}.

Continuing the analysis of the action \eqref{sgrav1}, the next term is the dilaton-curvature coupling, which arises from the CS term. The last term is just a total derivative. This term, just like the EH one, could only contribute to the field equations at the boundary if this boundary is non-trivial. 

Clearly, the action \eqref{sgrav1} is, up to boundary terms, manifestly gauge invariant under $SO(2)$ rotations in the tangent space,
\begin{eqnarray}
    \delta\Phi&=&0\;,\nonumber\\
    \delta e^a&=&\alpha^a_{\phantom{a}b}e^b\;,\nonumber\\
    \delta \omega^{ab}&=&d\alpha^{ab}\;.\label{gt3}
\end{eqnarray}
Thus, the action \eqref{sgrav1} is a genuine two-dimensional Euclidean gravity in the FOF, coupled to a scalar field. We point out that transformations \eqref{gt3} are directly obtained from \eqref{gt2}, \eqref{map1} and the identification $\alpha(x)=\epsilon_{ab}\alpha^{ab}(X)$, see \cite{Sobreiro:2021leg,Bittencourt:2025laf}. We also stress that action \eqref{sgrav1} exhibits diffeomorphism symmetry, which is implicit in the form notation. 

In the next section, we will show that this action can be transmuted into JT-type gravities. For now, let us focus on the dynamics of the action \eqref{sgrav1}. The field equation for $\Phi$ is just
\begin{equation}
    d\ast d\Phi+\Lambda\kappa\epsilon_{ab}R^{ab}=0\;,\label{feq0a}
\end{equation}
indicating that the scalar field carries some dynamics. It also suggests that the scalar field non-trivially influences the geometry of the manifold. However, the spin connection equation gives
\begin{equation}
    d\Phi=0\;,\label{feq0b}
\end{equation}
So, the dilaton solution is simply an arbitrary constant field configuration, possibly zero. Furthermore, combining equations \eqref{feq0a} and \eqref{feq0b} one easily finds that the manifold is locally flat:
\begin{equation}
    R^{ab}=0\;.\label{feq0c}
\end{equation}

It remains to derive the equation of the zweibein. The only nontrivial term remaining in the action \eqref{sgrav1} that depends on the zweibein is the area term. Therefore, the final equation simply corresponds to the minimization of the area of the manifold (Plateau problem), just as two-dimensional Mardones-Zanelli gravity \cite{Mardones:1990qc,Zanelli:2005sa}. The solution here is thus a flat space with minimal area. Of course, the specific manifold depends on the boundary conditions. Typical noncompact examples are planes, cylinders, and disks. In the case of compact manifolds, the flat torus is the standard solution. See Figure \ref{fig:cilindrodiscotorus}, in Section \ref{GB}, for examples. In fact, we will return to the discussion of these surfaces in Section \ref{GB}.

\section{Euclidean Jackiw-Teitelboim gravity}\label{JT}

In this section, we demonstrate that the action \eqref{sgrav1} can be related to the JT model \cite{Teitelboim:1983ux,Jackiw:1984je,Mertens:2022irh}, resulting in two types of manifolds: hyperbolic and spherical.

\subsection{Hyperbolic solution}\label{TRAC}

The first way to obtain JT gravity is by performing the curvature shift
\begin{equation}
    R^{ab}\longrightarrow R^{ab}+\frac{\Lambda^2}{2}e^ae^b\;,\label{shift1a}
\end{equation}
enforcing the usual JT term to appear directly in the gravity action \eqref{sgrav1}. However, before performing this shift properly, let us define it a bit more rigorously. In fact, the shift \eqref{shift1a} is actually generated by a shift in the spin connection, namely,
\begin{equation}
\omega^{ab}\longrightarrow\omega^{ab}+\frac{\Lambda^2}{2}X^{ab}\;,\label{shift1b}
\end{equation}
with the local condition
\begin{equation}
    dX^{ab}=e^ae^b\;,\label{cond1a}
\end{equation}
meaning that the field $X$ must be a non-conservative field with a canonical dimension of $[X]=-1$. For further reference, we name the condition \eqref{cond1a} \emph{shift constraint}. Thus, once the zweibein $e$ describing the manifold is chosen (note that there are infinite possibilities due to the $SO(2)$ gauge symmetry and diffeomorphisms), the field $X$ can be determined. Integrating equation \eqref{cond1a} and applying Stokes’ theorem gives the global condition
\begin{equation}
    \oint_C X^{ab}=\epsilon^{ab}\int_S d^2x\;.\label{cond1b}
\end{equation}
Thus, the circulation of $X$ is essentially determined by the area $S$ inside the closed curve of circulation $C=\partial S$. Typically, $S$ is of free choice; for instance, one can set $S\equiv M^2$, if $\partial M^2\ne0$.

Another point of importance concerning equation \eqref{cond1a} is that it is not gauge covariant unless $dX$ is covariant, which is actually not true. In fact, the shift \eqref{shift1b} suggests that $X$ is a gauge invariant field. As a consequence, gauge symmetry is broken by the constraint \eqref{cond1a}; it may be seen, thus, as a gauge fixing for the model. In fact, together with the shift \eqref{shift1b}, it can be incorporated into the action \eqref{sgrav1} with a suitable 0-form Lagrange multiplier\footnote{The Lagrange multiplier carries canonical dimension 3.} $\lambda_{ab}$ by means of the addition of the action
\begin{equation}
    S_\lambda=\int\lambda_{ab}\left(dX^{ab}-e^ae^b\right)\;.\label{gf1a}
\end{equation}
This term clearly breaks gauge symmetry, assuming that $\lambda$ is either gauge-invariant or covariant. 

Therefore, by employing the shift \eqref{shift1b} in the action \eqref{sgrav1} and the shift constraint via the action \eqref{gf1a}, we have the full action
\begin{eqnarray}
    S_{JT\lambda1}&=&\int\left\{-\frac{1}{\Lambda}d\Phi\ast d\Phi-\frac{1}{8\pi G}\epsilon_{ab}\left[R^{ab}+\frac{\Lambda^2}{2}\left(dX^{ab}-e^ae^b\right)\right]+2\kappa\Phi\epsilon_{ab}\left(R^{ab}+\frac{\Lambda^2}{2}dX^{ab}\right)+\right.\nonumber\\
    &+&\left.\lambda_{ab}\left(dX^{ab}-e^ae^b\right)+\kappa\epsilon_{ab}d\left(\Phi\omega^{ab}+\frac{\Lambda^2}{2}\Phi X^{ab}\right)\right\}\;.\label{sjt1b}
\end{eqnarray}
Let us derive the field equations. The equation for the Lagrange multiplier naturally imposes the shift constraint \eqref{cond1a}. The equation of the spin connection remains unchanged, equation \eqref{feq0b}. Thus, $\Phi$ is again undetermined. It is constant, but not necessarily zero. However, nothing prevents us from choosing $\Phi=0$. The equation for $X$, provides (already employing \eqref{feq0b})
\begin{equation}
    d\lambda^{ab}=0\;.
\end{equation}
implying that $\lambda$ has a constant configuration. 

Computing the field equation for $\Phi$, already applying \eqref{feq0b} and the equation for $\lambda$ (the shift constraint), provides the equation for hyperbolic manifolds,
\begin{equation}
    R^{ab}+\frac{\Lambda^2}{2}e^ae^b=0\;,\label{feq1b}
\end{equation}
which corresponds to the JT solution for negative constant curvature in Euclidean spaces.

Now, the zweibein field equation, already neglecting the kinematical term due to \eqref{feq0b}, gives
\begin{equation}
    \lambda_{ab}=\frac{\Lambda^2}{16\pi G}\epsilon_{ab}\;.\label{lambda1}
\end{equation}
Remarkably, due to the constraint \eqref{id1}, $\lambda$ can be written in terms of the cosmological constant:
\begin{equation}
\lambda_{ab}=\Lambda^3\epsilon_{ab}\;.\label{lambda2}
\end{equation}
This is actually quite an interesting result because a Lagrange multiplier functions as a force that enforces the system to satisfy the constraint. It turns out that the cosmological constant plays the role of that force. In essence, the cosmological constant is the force that bends the manifold. 

The solution we found so far is just a trivial dilatonic vacuum and a manifold with negative constant curvature, \emph{i.e.}, a hyperbolic surface. In two-dimensional Euclidean spaces, such manifolds, if compact, are the connected $n$-holed tori \cite{Stillwell1992,Anderson2005} - see Figure \ref{fig:ntorus} and Section \ref{GB} for further discussion.

\subsection{Spherical solution}\label{SPH}

The second way to achieve a JT solution is to perform the same shift \eqref{shift1a}, but with a different sign at the cosmological constant factor,
\begin{equation}
    R^{ab}\longrightarrow R^{ab}-\frac{\Lambda^2}{2}e^ae^b\;.\label{shift2a}
\end{equation}
Just like the previous case, the shift \eqref{shift2a} demands a more fundamental shift, here described by
\begin{equation}
\omega^{ab}\longrightarrow\omega^{ab}-\frac{\Lambda^2}{2}X^{ab}\;.\label{shift2b}
\end{equation}
Therefore, the shift constraint \eqref{cond1a} (or equivalently, \eqref{cond1b}) remains valid. In fact, the whole discussion about the shift constraint in the hyperbolic case remains valid here. Thence, performing the shift \eqref{shift2b} in the gravity action \eqref{sgrav1} and adding the gauge fixing term \eqref{gf1a} to the action results in the full action
\begin{eqnarray}
    S_{JT\lambda2}&=&\int\left\{-\frac{1}{\Lambda}d\Phi\ast d\Phi-\frac{1}{8\pi G}\epsilon_{ab}\left[R^{ab}-\frac{\Lambda^2}{2}\left(dX^{ab}+e^ae^b\right)\right]+2\kappa\Phi\epsilon_{ab}\left(R^{ab}-\frac{\Lambda^2}{2}dX^{ab}\right)+\right.\nonumber\\
    &+&\left.\lambda_{ab}\left(dX^{ab}-e^ae^b\right)+\kappa\epsilon_{ab}d\left(\Phi\omega^{ab}-\frac{\Lambda^2}{2}\Phi X^{ab}\right)\right\}\;.\label{sjt2a}
\end{eqnarray}
Let us compute the field equations. Equation \eqref{feq0b} remains valid, stating that $\Phi$ is a constant field again. The equation of $\lambda$ simply enforces the shift constraint \eqref{cond1a}. Again, the equation of the field $X$ gives that $d\lambda=0$, due to \eqref{feq0b}. Thus, $\lambda$ is also a constant field. The field equation for the dilaton, already employing the equations of $\omega$ and $\lambda$, gives the JT solution for a spherical manifold (constant positive curvature),
\begin{equation}
    R^{ab}-\frac{\Lambda^2}{2}e^ae^b=0\;.\label{feq2a}
\end{equation}

Finally, the equation for the zweibein again provides a fixed value for the Lagrange multiplier, see Equations \eqref{lambda1} and \eqref{lambda2}. And the dilaton remains undetermined, with the freedom of setting it to vanish.

\subsection{Extra remarks about the shift constraint}

To conclude this Section, we remark that the shift constraint \eqref{cond1a} can also be visualized as the imposition of a background. Therefore, the gauge fixing is actually a choice of an appropriate background. The only distinction between the hyperbolic and the spherical cases lies in the relative sign of the background. Essentially, the shifts \eqref{shift1b} and \eqref{shift2b}, together with the constraint \eqref{cond1a}, impose the background to be hyperbolic or spherical. Moreover, since in two dimensions the EH term is topological, there is no room to derive Einstein-like equations that would provide dynamics around the background.

In fact, the shifts \eqref{shift1b} and \eqref{shift2b} are simply stating that $\sim\pm X$ is the spin connection background in each case. The shift constraint \eqref{cond1a} states that $X$ must be the field that generates the specific (hyperbolic or spherical) background curvature. Moreover, it is known that the spin connection carries a specific term depending on the zweibein. Therefore, it is only natural that the shift constraint is written in terms of the zweibein. This is even evidenced by the fact that torsion is zero in our solutions.

One important piece of evidence that the model is not changed and the shift constraint, together with the shifts, is actually a background choice is the computation of the saddle point of actions \eqref{sgrav1}, \eqref{sjt1b}, and \eqref{sjt2a}. They all converge to the same value, namely
\begin{equation}
S_{2grav}\Big|_{on-shell}=S_{JT\lambda1}\Big|_{on-shell}=S_{JT\lambda2}\Big|_{on-shell}=\frac{\Lambda^2}{16\pi G}\mathcal{A}\;,\label{saddle}
\end{equation}
where $\mathcal{A}$ is the area of the manifold (or, at least, a specific piece of the manifold, if the manifold is not completely integrable).

It is also interesting to notice that the JT term could emerge directly from the mapping by considering an electromagnetic background field. Right from the action \eqref{sqed5}, one could consider the shift $F\longrightarrow F\pm\Theta/2$. Or, equivalently, a shift on the electromagnetic field $a\longrightarrow a\pm M$ with $M$ being fixed by $dM=\Theta/2$. In this way, the target gravity action would already contain the usual JT term $\Phi(R\pm\Lambda^2/2)$. Hence, the shift we proposed is equivalent to the existence of a background electromagnetic field in the original MCS theory.

Another comment is that the shift constraint \eqref{cond1a} induces a relation between the spin connection and the field $X$. This relation is actually generated by the cohomology of the nilpotent exterior derivative $d$, the well-known de Rham cohomology \cite{Nakahara:2003nw}. In fact, the constraint \eqref{cond1a} together with the curvature equations \eqref{feq1b} (negative sign) and \eqref{feq2a} (positive sign) can be combined to deduce the on-shell relation
\begin{equation}
    \omega^{ab}=\pm\frac{\Lambda^2}{2}X^{ab}+Y^{ab}\;,\label{xy1a}
\end{equation}
with $Y$ being a closed 1-form with dimension $[Y]=1$,
\begin{equation}
    dY^{ab}=0\;.\label{y1a}
\end{equation}
Moreover, since $X$ is gauge invariant, $Y$ must transform like a connection in order to preserve the gauge structure of relation \eqref{xy1a}, namely
\begin{equation}
    \delta Y^{ab}=d\alpha^{ab}\;.\label{y1b}
\end{equation}
Therefore, the gauge invariant field $X$ is essentially a difference between two connections, $X=\mp 2(\omega-Y)/\Lambda^2$, \emph{i.e.}, the difference between the spin connection and an arbitrary closed 1-form.

Condition \eqref{y1a} is actually a traditional cohomology problem whose solution is simply given by its non-trivial and trivial parts in de Rham cohomology, namely
\begin{equation}
    Y^{ab}=Z^{ab}+dN^{ab}\;,\label{yzn1a}
\end{equation}
with $Z$ being a closed non-exact 1-form with dimension $[Z]=1$,
\begin{equation}
    dZ^{ab}=0\;\Big|\;Z^{ab}\ne d(\mathrm{something})\;,\label{z1a}
\end{equation} 
while $N$ is an arbitrary 0-form (and hence cannot be written as an exact quantity) with dimension $[N]=0$. The gauge transformations of $Z$ and $N$ can be inferred from \eqref{y1b} and \eqref{yzn1a}. Clearly, $Z$ can be chosen to be a gauge connection while $N$ is set to be gauge invariant\footnote{Alternatively, there is also the possibility to choose $Z$ to be a gauge invariant field while $N$ would transform in a non usual way by $\delta N=\alpha$, \emph{i.e.}, $N$ would \emph{translate} under $SO(2)$ gauge transformations. Such transformation indicates that $N$ could be interpreted as a field of angular nature. In this case, the relation \eqref{yzn1a} could be seen as a gauge transformation of the field $Y$ while $N$ would be the gauge parameter.},
\begin{eqnarray}
    \delta Z^{ab}&=&d\alpha^{ab}\;,\nonumber\\
    \delta N^{ab}&=&0\;.\label{gt5}
\end{eqnarray}

The determination of the spin connection proceeds as follows. Once the zweibein describing the manifold is chosen, one can find any $X$ satisfying the shift constraint  \eqref{cond1a}, a closed non-exact field $Z$, satisfying \eqref{z1a}, and any gauge-invariant 0-form $N$. Then, using  relations \eqref{xy1a} and \eqref{yzn1a}, the spin connection is completely determined. The same reasoning works to find $X$ by solving equation \eqref{feq1b} (or \eqref{feq2a}) for the spin connection. In essence, the spin connection and the field $X$ differ only by any non-trivial field $Z$ plus an exact field. Moreover, $Z$ and $N$ take no part in the solutions we found for JT gravity; see Equations \eqref{feq1b} and \eqref{feq2a}. Alternatively, $X$ and $\omega$ can be deduced independently from equations \eqref{cond1a} and \eqref{feq1b} (or \eqref{feq2a}), respectively. Then, $Z$ and $N$ can be inferred by comparing $X$ and $\omega$ via relation \eqref{xy1a}.

It is also important to stress that $Y$ must be non-vanishing because of the gauge transformations of $\omega$, $Y$ (see \eqref{gt3} and \eqref{y1a}), and the gauge-invariant field $X$. Otherwise, relation \eqref{xy1a} would not be gauge covariant. Thus, if $Z\ne0$, then $dN$ can be set to vanish. On the other hand, since $N$ is gauge-invariant, $Z$ cannot be set to zero. Therefore, from now on, we are allowed to fix $N=0$ with no loss of generality. As an alternative, if we do not care about gauge covariance (which is a valid choice since the presence of $X$ in the action already breaks gauge symmetry), we can set $Z=N=0$ and work with a gauge-invariant spin connection. However, we will not follow that approach.

Now that we have all necessary fields in the gravity model, we display the dimensions and form ranks of all fields and parameters in Table \ref{table0b}.
\begin{table}[ht]
\centering
\begin{tabular}{|c|c|c|c|c|c|c|c|c|c|c|c|c|}
	\hline 
Fields & $e^a$ & $\omega^{ab}$ & $\Phi$ & $X$ & $Y$ & $Z$ & $N$ & $\lambda$ & $G$ & $\Lambda$ & $\kappa$\\
	\hline 
Dimension & $0$ & $1$ & $1$ & $-1$ & $1$ & $1$ & $0$ & $3$ & $-1$ & $1$ & $0$\\ 
Rank & $1$ & $1$ & $0$ & $1$ & $1$ & $1$ & $0$ & $0$ & $0$ & $0$ & $0$\\
\hline 
\end{tabular}
\caption{Canonical dimensions and form ranks of the fields and parameters in the gravity theory.}
\label{table0b}
\end{table}
The set of fundamental fields defined on $M^2$ is thus $\mathcal{F}=\{e,\omega,X,Z,\Phi,\lambda\}$, interlinked by the field equations
\begin{eqnarray}
    d\Phi&=&dZ^{ab}\;\;=0\;,\nonumber\\
    \lambda_{ab}&=&\Lambda^3\epsilon_{ab}\;,\nonumber\\
    R^{ab}&=&\pm\frac{\Lambda^2}{2}e^ae^b\;,\nonumber\\
    \omega^{ab}&=&\pm\frac{\Lambda^2}{2}X^{ab}+Z^{ab}\;.\label{fullfeq}
\end{eqnarray}
The third equation in \eqref{fullfeq} establishes that, among $\omega$, $X$, and $Z$, we can neglect one of them in favor of the other two. The space of fundamental fields is then reduced, for instance, to $\mathcal{F}_o=\{e,X,Z,\Phi,\lambda\}$, where we have chosen to omit the spin connection. 

We also remark that in all three cases (flat, hyperbolic, and spherical), the action has no torsional terms, with the torsion 2-form being defined as the covariant derivative of the zweibein, $T^a=De^a=de^a+\omega^a_{\phantom{a}b}e^b$. Nevertheless, torsion could be generated depending on the gravity-matter coupling. In dimensions greater than two, torsion and curvature are related by the Bianchi identity $DT\sim Re$ \cite{Mardones:1990qc,Zanelli:2005sa}. Thence, $T$ could be inferred from this relation. However, in two dimensions, this Bianchi identity vanishes identically, leaving torsion to be completely free\footnote{The same effect occurs in pure the Mardones-Zannelli gravity \cite{Mardones:1990qc,Zanelli:2005sa}.}; it depends solely on the choices we make in the solutions for $e$ and $\omega$. Therefore, one can simply set torsion to a fixed value and work one extra equation together with \eqref{fullfeq}. For instance, choosing torsion to be zero and using the last relation in Equations \eqref{fullfeq}, it is found that $Z$ (or $X$) is no longer independent,
\begin{equation}
    Z^a_{\phantom{a}b}e^b=-de^a\mp\frac{\Lambda^2}{2}X^a_{\phantom{a}b}e^b\;,\label{torsion0a}
\end{equation}
The set of independent fields can be thus even reduced to $\mathcal{F}_0=\{e,Z,\Phi,\lambda\}$.

\section{Gauss-Bonnet theorem: computing sizes}\label{GB}

The GB theorem\footnote{In the discussion provided here of the GB theorem, we restrict ourselves to two-dimensional Riemannian manifolds. For generalizations to any dimension, see \cite{Harder1971,Gromov1982}.} \cite{Stillwell1992,Nakahara:2003nw} is a powerful tool in the study of manifolds. The standard GB theorem for a connected and compact manifold, say $M^2$, states that \cite{Stillwell1992,Nakahara:2003nw},
\begin{equation}
\int_{M^2}\epsilon_{ab}R^{ab}=4\pi\chi\;,\label{gb1a}
\end{equation}
with $\chi$ being the Euler characteristic of the manifold. It is known that the GB theorem \eqref{gb1a} generalizes for complete, finitely connected, and noncompact manifolds \cite{Rosenberg1982,CohnVossen1935,Huber1957,Kellerhals2001} if the integral is absolutely integrable and the total area is finite. Nevertheless, we mostly confine ourselves to the compact surface cases. 

The formula \eqref{gb1a} can be employed to select more accurately the possible manifolds of our solutions. And, in some cases, it can be employed to determine their sizes, based on their Euler characteristics.

\subsection{Flat surfaces}

The solution discussed in Section \ref{GRAV} is a locally flat manifold of minimal area. Considering the GB theorem for compact surfaces of vanishing curvature, the GB theorem in the form \eqref{gb1a} restricts the pool of solutions to manifolds with zero Euler characteristic. It turns out that the manifold satisfying all requirements of the GB theorem is the torus with only one handle \cite{Stillwell1992}.

It turns out that formula \eqref{gb1a} can also be applied to noncompact surfaces such as the cylinder \cite{Stillwell1992,Massey1962}. In fact, any manifold topologically equivalent to a cylinder or a torus, while maintaining the flatness of the surface, can be considered as a solution. For example, a disk with a hole in it satisfies the criteria. Notice that the cylinder and the holed disk are topologically equivalent; both are obtained from the continuous deformation of a sphere with two holes. 

\begin{figure}[ht]
    \centering
    \includegraphics[scale=0.3]{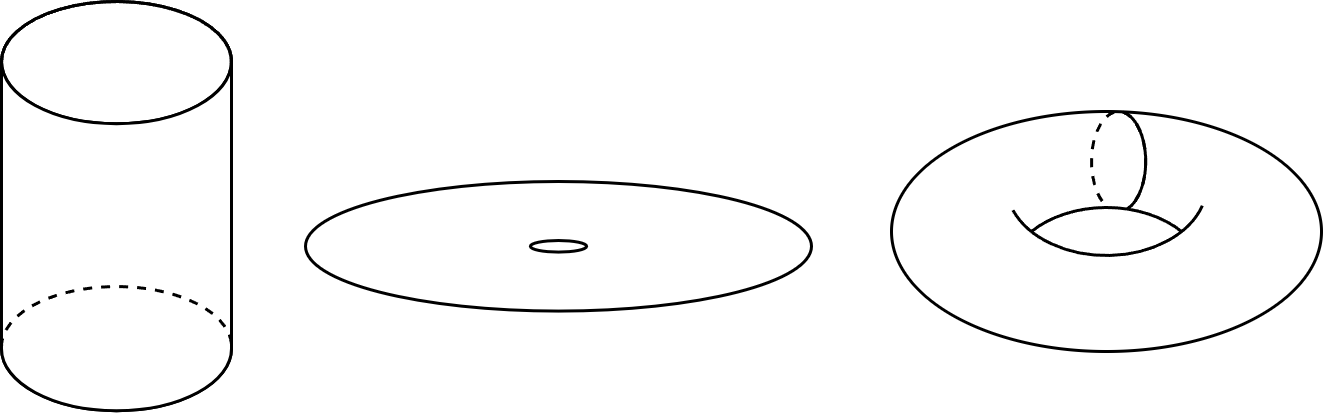}
    \caption{Flat surfaces with vanishing Euler characteristic. From left to right: The cylinder; the flat disk with a hole; the flat torus. Among these manifolds, only the flat torus is compact.}
    \label{fig:cilindrodiscotorus}
   \centering
\end{figure}

Unfortunately, the size of the surfaces with vanishing curvature cannot be computed from the GB theorem. Nevertheless, the theorem can be used to select the possible solutions to be in the set of surfaces with $\chi=0$. Some examples of flat surfaces with vanishing Euler characteristics are displayed in Figure \ref{fig:cilindrodiscotorus}.

\subsection{Hyperbolic surfaces}

\begin{figure}[ht]
    \centering
    \includegraphics[scale=0.35]{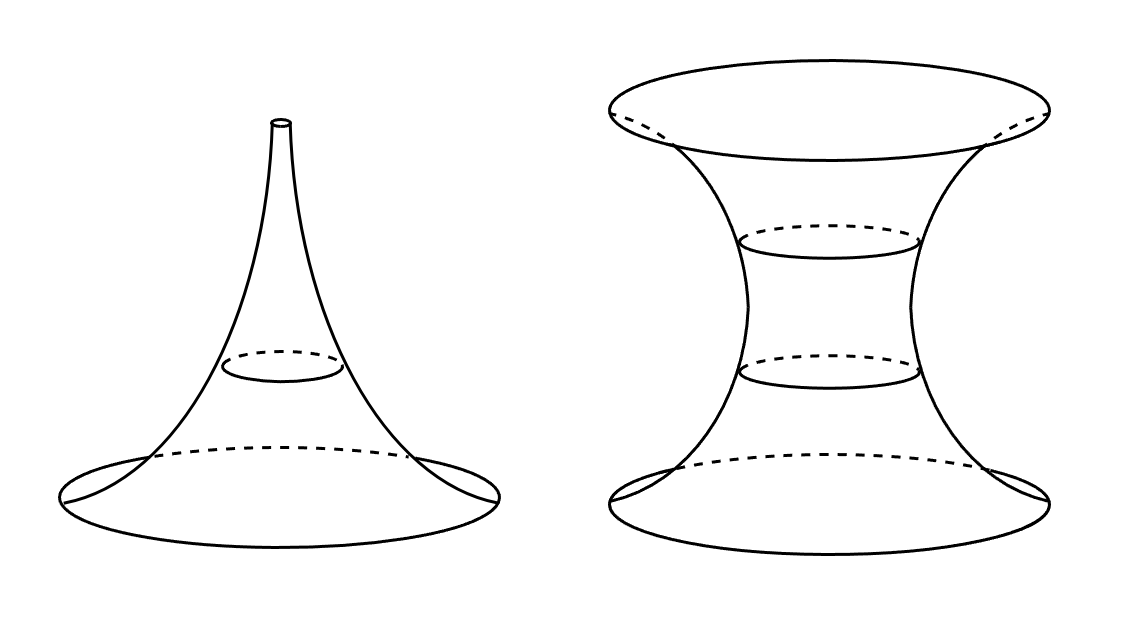}
    \caption{Noncompact hyperbolic manifolds. From left to right: The tractroid; and the catenoid. These manifolds have vanishing Euler characteristics.}
    \label{fig:tractroide}
   \centering
\end{figure}

The solution found in Section \ref{TRAC} is a manifold of negative constant curvature, \emph{i.e.}, any hyperbolic surface \cite{Stillwell1992,Anderson2005}. Typical noncompact examples are the tractroid (pseudo-sphere) and the catenoid (See Figure \eqref{fig:tractroide}), which is homeomorphic to the cylinder, and thus has a vanishing Euler characteristic as well \cite{Stillwell1992}. Therefore, we cannot employ the GB theorem \eqref{gb1a} to compute its size. For compact hyperbolic surfaces, the simplest manifolds with negative curvature are essentially those which are equivalent to the $n$ linearly connected tori ($n$-holed torus), where $n$ is the number of handles (related to its genus) and the corresponding Euler characteristic is given by $\chi=2-2n$, with $n>1$ (The case $n=1$ is the simple torus with only one handle and vanishing curvature - see the previous Section) \cite{Stillwell1992,Anderson2005}. Clearly, for hyperbolic surfaces, $\chi$ must be negative. Of course, the tori can be nonlinearly glued, but the resulting genus would be bigger than $n$, as we will approach in a following example. Thus, putting the solution \eqref{feq1b} in the GB theorem \eqref{gb1a}, we find\footnote{We remark that it is commonly known the validity of the formula
\begin{equation}
    \mathcal{A}_h=-2\pi\chi\;,
\end{equation}
for the area of compact surfaces \cite{Stillwell1992}. At a first glance, this formula seems to not be compatible with formula \eqref{Ah0}. Nevertheless, this formula is deduced for surfaces with normalized curvature, which is not the case here because of the presence of the cosmological constant.}
\begin{equation}
\mathcal{A}_h=\frac{8\pi(n-1)}{\Lambda^2}\;,\label{gb1b}
\end{equation}
with $\mathcal{A}_h$ being the area of the $n$-holed torus. Substituting the value of the cosmological constant computed in expression \eqref{num0} in formula \eqref{gb1b}, gives
\begin{equation}
    \mathcal{A}_h=3,102.8(n-1)eV^{-2}\;,\label{Ah0}
\end{equation}
which, in SI units, corresponds to 
\begin{equation}
    \mathcal{A}_h=120.42(n-1)\mathrm{\mu m}^2\;.\label{Ahn}
\end{equation}

\begin{figure}[ht]
    \centering
    \includegraphics[scale=0.3]{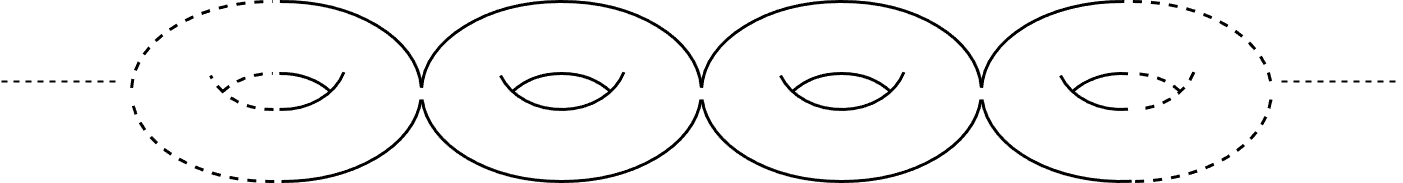}
    \caption{The $n$-holed torus. A compact hyperbolical surface with non-vanishing Euler characteristics ($\chi=2-2n$). To compute the size of this manifold in an easy way, the intersections of the tori are made very small.}
    \label{fig:ntorus}
   \centering
\end{figure}

Let us consider a first example, for instance (see Figure \ref{fig:ntorus}), $n$ identical tori, linearly connected, with two equal radii $2b_h$ and $b_h$, and making the interface between each tori a circle of radius $c<<b_h$ (In this way, we characterized the manifold by only one parameter). This is the well-known $n$-holed torus. We can determine the radius $b_h$ from formula \eqref{Ahn}:
\begin{equation}
    b_h=1.23\sqrt{\frac{(n-1)}{n}}\mu m\;.
\end{equation}
The minimum possible value is thus $n=2$, providing $b_h=0.87\mu m$. As $n\longrightarrow\infty$, the area stabilizes at the maximum value of $b_h=1.23\mu m$. We conclude that, although the area diverges with $n$, the radii of the identical $n$-holed torus do not.

\begin{figure}[ht]
    \centering
    \includegraphics[scale=0.25]{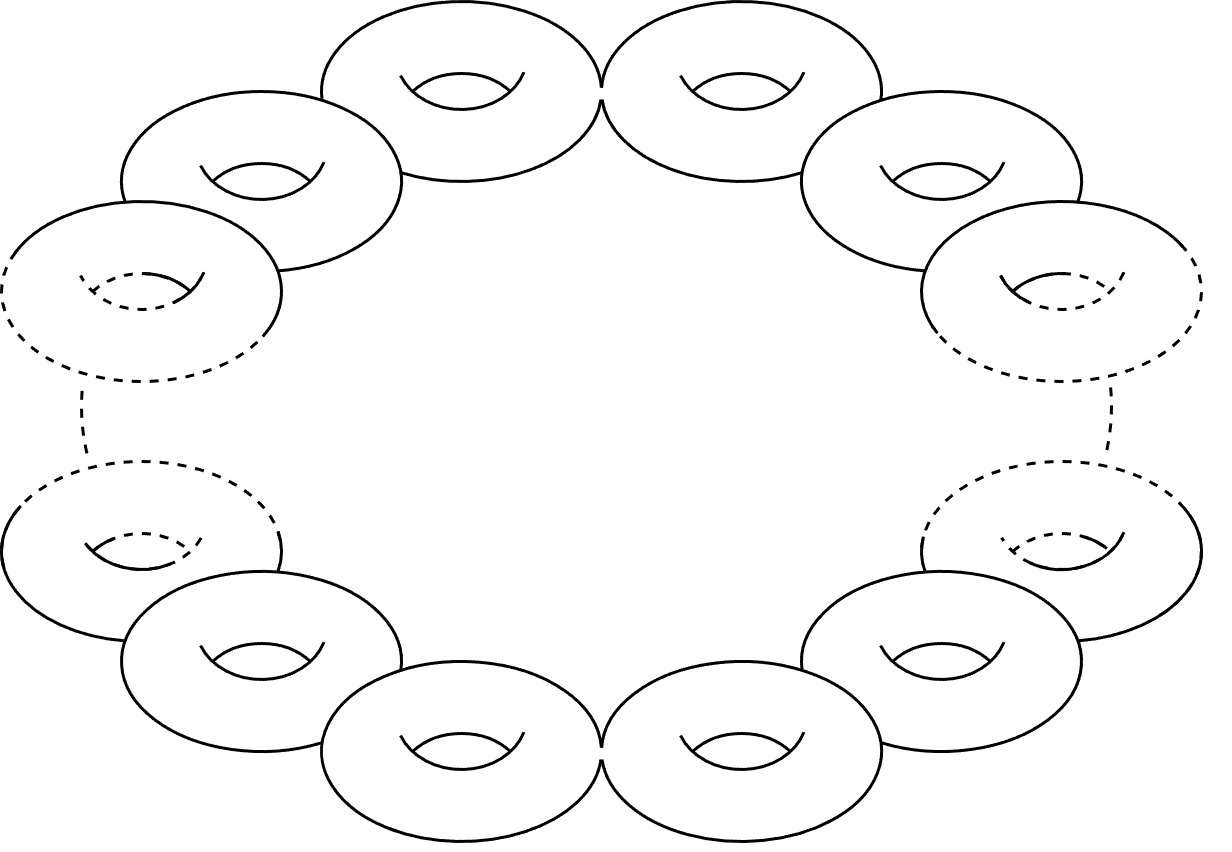}
    \caption{The closed $n$-holed torus. They are linearly connected and the tori of the edges are also connected, forming a ring of tori. The big hole in the middle increases its Euler characteristics by one ($\chi=-2n$). Just like the open linearly connected $n$-holed torus, the intersections of the tori are made very small in order to compute the size of this manifold in an easy way.}
    \label{fig:aneltorus}
   \centering
\end{figure}

Just to illustrate another case, where the genus changes, we consider now the same $n$-holed torus of the previous example. However, we glue the first and last tori together, to form ring composed of $n$ tori, see Figure \ref{fig:aneltorus}. In this case, the genus increases by one, $\chi=2-2(n+1)=-2n$. Thence,
\begin{equation}
\mathcal{A}_h=\frac{8\pi n}{\Lambda^2}\;,\label{gb1c}
\end{equation}
implying that $b_h$ no longer depends on $n$ and is fixed by the value $b_h=1.23\mu m$.

These results are quite remarkable because it essentially says that the sizes of some surfaces in the present gravity model are ultimately determined by the electric charge in the original model, \emph{i.e.}, the manifold scale in one theory is linked to the electric charge in another theory.

\subsection{Spherical surfaces}

In Section \ref{SPH}, another sort of solution was derived. Essentially, manifolds with constant positive curvature (see \eqref{feq2a}). Traditional noncompact examples include spherical caps and curved disks. For compact solutions, we have the usual sphere itself.

\begin{figure}[ht]
    \centering
    \includegraphics[scale=0.35]{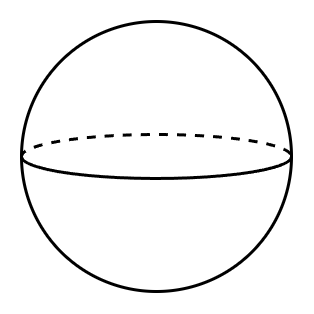}
    \caption{The sphere. A compact manifold with positive curvature and non-trivial Euler characteristics ($\chi=2$).}
    \label{fig:esfera}
   \centering
\end{figure}

Among all possible surfaces, we consider the standard sphere as an example (see Figure \ref{fig:esfera}). Therefore, direct integration of formula \eqref{feq2a} and the use of expression \eqref{gb1a} imply the relation $\Lambda^2\mathcal{A}_s=4\pi\chi$, with $\mathcal{A}_s$ being the area of the sphere. For the sphere, the Euler characteristic is given by $\chi=2$. As a consequence, its area is fixed by $\mathcal{A}_s=8\pi/\Lambda^2$ and the radius $b_s$ of the sphere is thus determined by the simple formula
\begin{equation}
    b_s=\frac{\sqrt{2}}{\Lambda}\;.\label{rs1}
\end{equation}
Substituting the numerical value of the cosmological constant, as computed in expression \eqref{num0}, we obtain 
\begin{equation}
b_s=15.7eV^{-1}\;.\label{rs2}
\end{equation}
In SI units, the radius and the area of the sphere are $3.09\mu m$ and $119.98\mu m^2$, respectively. 

Moreover, it is possible to relate the radii $b_h$ and $b_s$ by combining their area formulas. For instance, in the case of the open linearly connected $n$-holed torus, we have
\begin{equation}
    \mathcal{A}_h=(n-1)\mathcal{A}_s\;.
\end{equation}
Therefore, considering as an example the system of $n$ identical tori from the previous Section, we get
\begin{equation}
    b_h=\sqrt{\frac{(n-1)}{2\pi n}}b_s\;.
\end{equation}
For the minimal ($n=2$) and extremal ($n\longrightarrow\infty$) cases we have a small range of values,
\begin{eqnarray}
    \lim_{n\rightarrow2}b_h&=&\frac{b_s}{2\sqrt{\pi}}\;,\nonumber\\
    \lim_{n\rightarrow\infty} b_h&=&\frac{b_s}{\sqrt{2\pi}}\;,
\end{eqnarray}
in full agreement with the first example in the previous Section.

\section{Conclusions and perspectives}\label{conc}

In this paper, we have shown that pure three-dimensional electrodynamics in the static limit corresponds to two-dimensional dilatonic gravity in Euclidean space. The Newtonian and cosmological constants are completely determined by the electric charge, as indicated in the relations \eqref{id0}. Furthermore, we were able to compute these constants, see Equation \eqref{num0}. Both of them are found to be several orders of magnitude larger than their four-dimensional experimental values, indicating a form of strong two-dimensional gravity.

We also presented three possible solutions for such a gravity model in Sections \ref{GRAV} and \ref{JT}, namely flat, hyperbolic, and spherical manifolds. These solutions essentially constitute the three possible situations of two-dimensional manifolds \cite{Stillwell1992}. In Section \ref{GB}, we employed the GB theorem to determine their sizes and some topological properties. Moreover, in all three solutions, the dilaton is a constant arbitrary field, which is consistent with its electromagnetic origin as a scalar potential. The main properties of the solutions discussed in these Sections are described in what follows:
\begin{itemize}
    \item {\bf Flat manifolds}: This solution is obtained from the action \eqref{sgrav1}, which originates directly from the mapping. In two dimensions, for various types of manifolds, the GB theorem, in formula \eqref{gb1a}, can be used to integrate the scalar curvature. Among these manifolds, since the curvature vanishes, we are restricted to manifolds with zero Euler characteristic. The possibilities we have inferred as examples are the cylinder (or, equivalently, a flat disk with a hole in it) and the flat torus (see Figure \ref{fig:cilindrodiscotorus} for the examples). The flat torus is, of course, the only compact surface in these examples. Unfortunately, the size of these manifolds cannot be determined by the GB theorem. 
    
    \item {\bf Hyperbolic manifold}: This solution is obtained by enforcing the shift constraint \eqref{cond1a}, related to the shift \eqref{shift1b}. The result is a JT-type gravity with manifolds of constant negative curvature. Among hyperbolic surfaces, the tractroid, the catenoid, and the $n$-holed torus (see Figure \ref{fig:tractroide} for the noncompact surfaces and Figures \ref{fig:ntorus} and \ref{fig:aneltorus} for the compact examples) can be taken as examples. Since the catenoid and the tractroid are topologically equivalent to the cylinder, they have vanishing Euler characteristics, preventing the use of the GB theorem \eqref{gb1a} to determine the sizes of these surfaces. 
    
    For the $n$-holed tori, we have applied the GB theorem to determine their sizes in two examples: the open linearly connected $n$-holed torus (see Figure \ref{fig:ntorus}) and the closed linearly connected $n$-holed torus (Figure \ref{fig:aneltorus}). The area of these manifolds depend on $n$ and $\Lambda$, as shown in expressions \eqref{gb1b} and \eqref{gb1c}. Explicit computations were performed by considering a specific size of each torus, taken to be identical and characterized only by one parameter, the radius of the inner circle of the transversal cross-section of one torus. It was found an incredibly small radius for these manifolds. For the open $n$-holed torus with $n=2$, for example, the radius is just $0.87\mu m$. As $n$ increases, the radius asymptotically stabilizes at the value of $1.23\mu m$. For the closed $n$-holed torus, the radius does not depend on $n$ and is fixed to the value of $1.23\mu m$. Naturally, these values are computed from the value of the cosmological constant, determined in \eqref{num0}, originally determined by the electric charge.
    
    \item {\bf Spherical manifolds}: This solution is obtained through the shift \eqref{shift2a} and the shift constraint \eqref{cond1a}, which is the same as for the hyperbolic solution. Constant positive curvature for JT gravity is obtained, namely \eqref{feq2a}. Therefore, the solution describes spherical manifolds. Among spherical surfaces, we take the usual sphere (Figure \ref{fig:esfera}), which is compact, as an example. Thence, GB theorem can be readily applied in order to determine the sphere radius in terms of the cosmological constant. Following the same steps developed in the $n$-holed tori examples, we were able to find that the radius of the sphere is remarkably small, valuing $3.09\mu m$.
\end{itemize}

A more technical comment is that the constraint \eqref{cond1a} (or the corresponding shifts of the curvature and spin connection) is, in practice, equivalent to choosing a background configuration. Particularly, hyperbolical or spherical ones. This background can only be chosen due to the presence of mass parameters in the action. Otherwise, a mass parameter would be required by hand in order to achieve JT gravity. Fortunately, we do not have any such mass parameter, but the specific parameter interpreted as a cosmological constant. Even if we did not have it, we recall that $G$ also carries mass dimension, allowing a different choice of curvature values. 

Ultimately, the background can also be defined before the mapping. As discussed in Section \ref{JT}, one could define a background field strength by the shift $F\longmapsto F\pm\Theta/2$ and the corresponding shift of the gauge potential $a$. The analogue of the shift constraint \eqref{cond1a} would appear in terms of the fields of the MCS theory. 

Another technical point is that the mapping \eqref{map1} is fundamentally based on the isomorphism $U(1)\longmapsto SO(2)$. Therefore, we end up in a Euclidean gravity model. Minkowskian gravity could be obtained, in principle, by performing a Wick rotation at the gravity isometry group, $SO(2)\longmapsto SO(1,1)$, inducing a Minkowskian signature in the manifold metric. Thus, instead of hyperbolic and spherical manifolds, the surfaces would be Anti-de Sitter and de Sitter spacetimes, respectively. Nevertheless, the detailed study of the Wick rotation in the mapping we defined and the theories we studied is beyond the scope of the present work.

A remarkable feature of the model is that the relationship between the gravitational constants and the electric charge, given in \eqref{id0}, indicates a possible duality because $G\sim1/e^2$. Thence, if on one hand we have electrodynamics as a perturbative system, on the other hand we have a JT gravity model with a strong coupling parameter. Alternatively, there is no obstruction in starting from gravity with any value of $G$ and map it in static MCS theory while maintaining the relation between $G$ and $e$. In that case, one could describe standard JT gravity in terms of MCS static electromagnetism with some exotic value of electric charge. Thus, for example, a black hole configuration in JT gravity could be described by an electromagnetic system with a specific kind of charged fields. Such ideas also suggest that our model also has the potential to be employed in condensed matter systems, as discussed in the Introduction.

A few additional remarks are in order. First, it is important to emphasize that we have used a map between pure static three-dimensional MCS electromagnetism and two-dimensional Euclidean gravity in the FOF. The electromagnetism side is a generic model with no characterized length scale (or energy scale). The only number we have considered is the fundamental charge of the electron $e$. On the gravity side, we found three possible solutions. Two of them have specific length scales of the order of $\mu m$ (micrometers). This feature originates from the fact that the curvatures of the exemplified surfaces are determined from the cosmological constant, while the latter is fixed by the electric charge. In essence, these two solutions are solutions of JT gravity with negative and positive curvatures.

In the same spirit, we can start from the gravity model we found and go back to electrodynamics at the static limit. Therefore, one might consider associating systems with characteristic lengths on the order of $\mu m$ with the properties of static MSC electrodynamics. Examples in nature of objects and living beings of such small size are: suspended fine dust; the mitochondrion; large viruses such as the \textit{Megavirus chilensis} \cite{Arslan2011} and the \textit{Acanthamoeba polyphaga Mimivirus} \cite{LaScola2003,Claverie2006}; some bacteria (\emph{e.g.}, the \textit{Escherichia coli}, also known as the E. coli, \cite{Poon2013}). Thus, perhaps, the surface of these systems could be modeled by MCS static theory in flat space.

Indeed, there are many perspectives for future investigations. We mention here some immediate possibilities. For instance, we can look at the four-dimensional case and its corresponding dimensional reduction (time and one space coordinate, for example). Naturally, one extra scalar field will emerge, and the dynamics can change. Another point is the inclusion of matter. Fermions could be considered in the same way we have considered them in \cite{Bittencourt:2025laf}, or one could simply include external source interactions. Of course, extra care will be demanded in the dimensional reduction of spinors \cite{Manton:1981es,Wetterich:1983ye,Wetterich:1984uc}. It could also be interesting to study matter with different electric charges. In this case, the violation of the equivalence principle is expected since only one type of charge can be absorbed in the gauge field. Moreover, inclusion of matter may be relevant for black hole analysis \cite{Cadoni:1994uf,Moitra:2019bub,Mertens:2022irh}. In terms of time evolution, one can simply decompose space and time of three-dimensional electrodynamics and map to gravity only the space coordinates. The expected result would be that of surfaces and the dilaton evolving in time as a separate coordinate of the manifold. The same idea can be generalized to four dimensions with only one space coordinate dimensional reduction - in that case, two scalar fields will appear, possibly evolving in time together with the manifold. One last perspective to address is that one could extend the discussion developed here to the case of general affine connections with non-metricity. Such work would require a separate and more detailed investigation, which goes beyond the scope of the present paper.

Finally, it is worth mentioning that, in two dimensions, the map \eqref{map1} is also consistent at the quantum level. Therefore, one can expect that the same feature to hold in the present case. Nevertheless, this study is beyond the scope of the present work.

\section*{Acknowledgements}

The authors are in debt to J. Zanelli, M. S. Guimarães, and G. Sadovski for valuable discussions and ideas. This study was financed in part by The Coordena\c c\~ao de Aperfei\c coamento de Pessoal de N\'ivel Superior - Brasil (CAPES) - Finance Code 001.

\bibliography{BIB}
\bibliographystyle{elsarticle-num}

\end{document}